\documentclass[sigconf]{acmart}

\usepackage{subcaption}

\AtBeginDocument{%
  \providecommand\BibTeX{{%
    \normalfont B\kern-0.5em{\scshape i\kern-0.25em b}\kern-0.8em\TeX}}}

    \newcommand\blfootnote[1]{%
  \begingroup
  \renewcommand\thefootnote{}\footnote{#1}%
  \addtocounter{footnote}{-1}%
  \endgroup
}

\setcopyright{acmcopyright}
\copyrightyear{2023}
\acmYear{2023}
\acmDOI{XXXXXXX.XXXXXXX}

\acmConference[MM '23]{Proceedings of the 30th ACM International Conference on Multimedia}{xx 03--05,
  2023}{xx, xx}
\acmPrice{15.00}
\acmISBN{978-1-4503-XXXX-X/18/06}

\begin{document}

\title[DCTM for Multimodal Engagement Estimation in Conversation]{DCTM: Dilated Convolutional Transformer Model for Multimodal Engagement Estimation in Conversation}


\author{Vu Ngoc Tu}
\affiliation{%
  \institution{Chonnam National University}
  \city{Gwangju}
  \country{South Korea}
  }
\email{tu369@jnu.ac.kr}

\author{Van Thong Huynh}
\affiliation{%
  \institution{Chonnam National University}
  \city{Gwangju}
  \country{South Korea}
}
\email{vthuynh@jnu.ac.kr}
\author{Hyung-Jeong Yang}
\affiliation{%
  \institution{Chonnam National University}
  \city{Gwangju}
  \country{South Korea}
}
\email{hjyang@jnu.ac.kr}

\author{M. Zaigham Zaheer}
\affiliation{%
 \institution{Mohamed bin Zayed University of AI}
 \streetaddress{Rono-Hills}
 \city{Masdar City}
 \state{Abu Dhabi}
 \country{UAE}
 }
\email{zaigham.zaheer@mbzuai.ac.ae}

\author{Shah Nawaz\textsuperscript{\textdagger}}
\affiliation{%
  \institution{Deutsches Elektronen-Synchrotron}
  \streetaddress{30 Shuangqing Rd}
  \city{Hamburg}
  % \state{Beijing Shi}
  \country{Germany}}
\email{shah.nawaz@desy.de}

\author{Karthik Nandakumar}
\affiliation{%
 \institution{Mohamed bin Zayed University of AI}
 \streetaddress{Rono-Hills}
 \city{Masdar City}
 \state{Abu Dhabi}
 \country{UAE}
 }
\email{karthik.nandakumar@mbzuai.ac.ae}

\author{Soo-Hyung Kim}
\authornote{Corresponding author}
\affiliation{%
  \institution{Chonnam National University}
  % \city{Gwangju}
  \country{South Korea}
}
\email{shkim@jnu.ac.kr}

\renewcommand{\shortauthors}{Tu, Thong, and Zaigham et al.}

\begin{abstract}
Conversational engagement estimation is posed as a regression problem, entailing the identification of the favorable attention and involvement of the participants in the conversation. This task arises as a crucial pursuit to gain insights into human's interaction dynamics and behavior patterns within a conversation. 
In this research, we introduce a dilated convolutional Transformer for modeling and estimating human engagement in the MULTIMEDIATE 2023 competition. Our proposed system surpasses the baseline models, exhibiting a noteworthy $7$\% improvement on test set and $4$\% on validation set. Moreover, we employ different modality fusion mechanism and show that for this type of data, a simple concatenated method with self-attention fusion gains the best performance.

\end{abstract}

\begin{CCSXML}
<ccs2012>
   <concept>
       <concept_id>10010147.10010178</concept_id>
       <concept_desc>Computing methodologies~Artificial intelligence</concept_desc>
       <concept_significance>500</concept_significance>
       </concept>
   %     <concept_id>10003120.10003121</concept_id>
   %     <concept_desc>Human-centered computing~Human computer interaction (HCI)</concept_desc>
   %     <concept_significance>500</concept_significance>
       </concept>
 </ccs2012>
\end{CCSXML}

\ccsdesc[500]{Computing methodologies~Artificial intelligence}
% \ccsdesc[500]{Human-centered computing~Human computer interaction (HCI)}

\keywords{engagement estimation, transformer, multimodal}

\maketitle
\section{Introduction}
\blfootnote{\textsuperscript{\textdagger}Current Affiliation: IMEC, Belgium}
Engagement is the process by which two (or more) participants
establish, maintain, and end their perceived connection to each
other during an interaction~\cite{sidner2002human}. 
Participant engagement stands as a key factor within the multifaceted dynamics of conversation, wielding significant influence over the quality and effectiveness of the interaction. However, while it is natural for human to discern the attentiveness of conversation counterparts, it remains a difficult task for a machine to apprehend~\cite{pellet2023multimodal}. Therefore, automatically estimating engagement degrees has became a primary challenge for both affective computing and group behavior analysis. The significance of this task has been increasingly recognized by researchers recently, primarily due to its widely ranging applications in various fields including education \cite{nomura2019estimation, karimah2022automatic}, human-computer interaction \cite{sidner2002human, oertel2011use,ooko2011estimating, oertel2013gaze}, social interaction \cite{song2012multimodal, sakaguchi2022estimation, rajagopalan2015play}, and healthcare \cite{lo2019engagement, graffigna2015measuring, zhang2022engagement}. 

To address this task, the Multimodal Group Behaviour Analysis for Artificial Mediation (MULTIMEDIATE 2023 \cite{muller2023multimediate}) challenge is organized. 
% Over the years, this prestigious competition has become a pivotal platform   for researchers in the domain of Artificial Mediation and Human behavior analysis. 
It encompasses two distinct tasks, namely engagement estimation and bodily behavior recognition in social interactions. The engagement estimation task involves quantifying participants' attention levels by leveraging various modalities. Simultaneously, bodily behavior recognition focuses on classifying specific behavior types through the analysis of human pose and facial expression data.  Within the scope of this paper, we prioritize our focus towards the Engagement Estimation task \cite{muller2023multimediate} .
% his challenge serves as a crucial platform for researchers specializing not only in Artificial Mediation but also in affective computing and human behavior analysis. It encompasses two distinct tasks, namely engagement estimation and bodily behavior recognition in social interactions, which are fundamental aspects of analyzing human social behavior. The engagement estimation task involves quantifying participants' attention levels by leveraging various modalities. Simultaneously, bodily behavior recognition focuses on classifying specific behavior types through the analysis of human pose and facial expression data. In this paper, we prioritize our attention on the Engagement Estimation task, as outlined in [15].

In some popular applications of engagement estimation system, such as education, we only require the head pose modality or facial expression to make a decision \cite{thong2019engagement, whitehill2014faces}. However, in the conversation context, human tend to interact with each other not only verbally but also with body language and facial expressions. Hence, to precisely determine whether a person is engaging in the conversation or not, we need to analyze the participants in terms of three attributes: their body attribute, head attribute and speech attribute \cite{pellet2023multimodal, song2012multimodal}. The scope of MULTIMEDIATE competition allows us to utilize all of these three attributes for developing an effective solution.

In recent years, the success of Transformer \cite{vaswani2017attention} model and its successors \cite{devlin2018bert, floridi2020gpt, dosovitskiy2020image, nie2022time} in different fields such as natural language processing \cite{devlin2018bert, floridi2020gpt}, time series analysis \cite{nie2022time}, and computer vision \cite{dosovitskiy2020image} gain the sequence model family a huge popularity and became the model of choice for different problems. In multimodal engagement estimation task, although there have been some study already carried out on the non-attention-based models like LSTM \cite{hochreiter1997long} and RNN \cite{rumelhart1986learning}, the attention-based model has not been thoroughly investigated. Moreover, we hypothesize that as the problem is multimodal with temporal information critical towards final prediction, an attention-based model such as transformers can be extremely effective.

To this end, we propose an architecture for engagement estimation that combines dilated convolution and transformers. We treat the modalities of the three attributes described previously as the signal and use them as the time-series-based data.

% \section{Related Works}

% Engagement Estimation emerges through out the years as one of the core problems of Human-Robot Interation (HRI) field. There is a trend in making the robot to be able to adapt their actions according to user behavior. On the other hands, there is a lack of number of research focusing on examing the factors of multimodal in human-human interaction. The early effort in trying to automatically detect the degree of human's engagement mostly come from the HRI field. Ooko et al (2011) \cite{ooko2011estimating} used the head tracking system to to analyze human's attentiveness in conversation. 
% Oertel et al. (2011)\cite{oertel2011use} used multimodal cues to predict the degree of
% group involvement during spontaneous conversation, extracting acoustic features and visual features. Oertel and Salvi (2013) \cite{oertel2013gaze} showed that engagement and involvement can be modelled by using gaze pattern. 
% Recently, Pellet et al \cite{pellet2023multimodal}  try to adress these problem by analyzing the different aspect of engagement in human-human conversation. In spite of having the promising result, this research still mainly utilize traditional Machine learning model to predict, which do not reflect the advancement of deep learning field.

\section{Methodology}
In this section, we introduce our proposed method for the estimation of continuous engagement.
\subsection{Problem statement}
The objective of engagement estimation is to predict frame-wise the degree of engagement from participant on the continuous scale ranging from 0 (lowest) to 1 (highest) from the input which is the multimodal signal. 
We formulate the engagement estimation as a regression problem on time-series data. 
\subsection{Dilated Convolutional Transformer model}
\begin{figure}
  \centering
   \includegraphics[width=.90\linewidth]{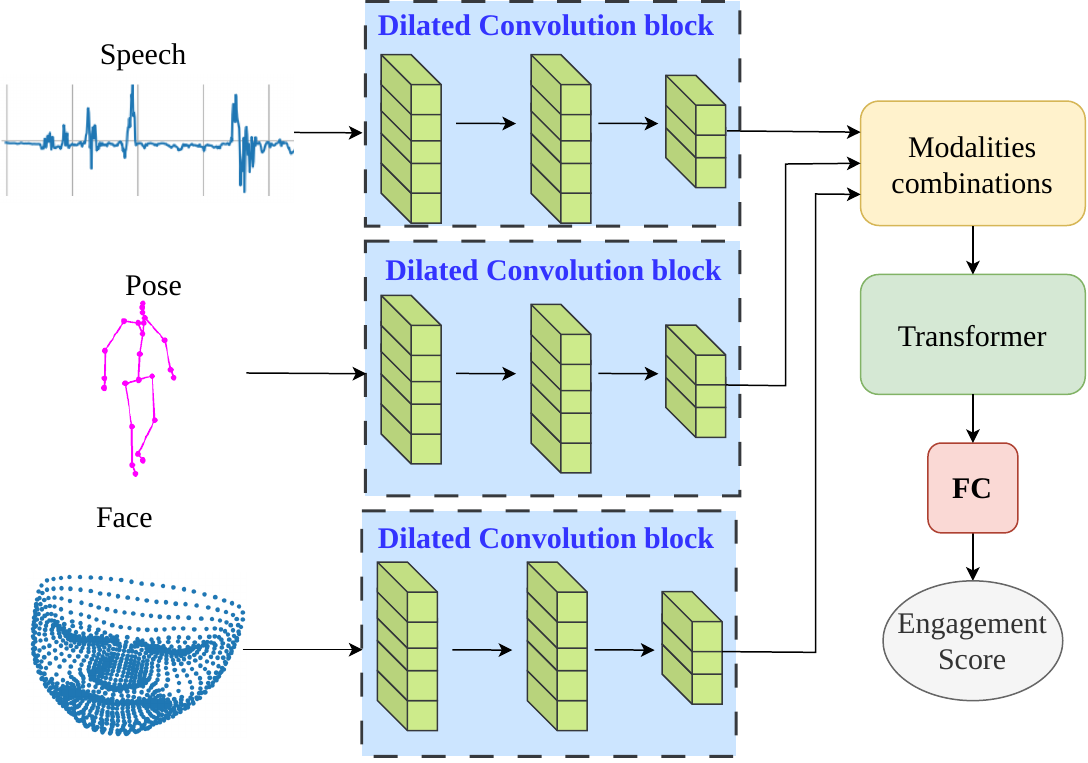}

   \caption{
   An overview of the engagement estimation model. The model leverages 3 modalities as input, passing them through convolutional layers before merging and sending them to the transformer for the desired outcome.}
   \label{fig:overal}
\end{figure}

Overall, our approach consists of three main components: the Long sequence feature extractor, the multiple modalities combination module and the Frame-wise regressor. Input of the model is the sequence of time-series data obtained from sliding window. The architecture is shown in Figure \ref{fig:overal}.
\subsubsection{Long Sequence Feature extraction}
During conversations, when a participant reaches a specific engagement state, the duration tends to be prolonged with minimal changes in the engagement score. Therefore, it becomes crucial to have a comprehensive model coverage that captures the overall trend and extracts global information from the sequence. However, the use of large convolutional filters can lead to overfitting, particularly due to the limited size of the available data.

To address this issue, we propose the utilization of dilated convolution, which allows us to enlarge the model's receptive field while preserving the input resolution throughout the network. Dilated Convolution, introduced by Holschneider et al. in $1990$ \cite{holschneider1990real}, Dilated Convolution has become a prominent method for signal processing. Since the first use in deep learning \cite{yu2015multi}, it has become one of the most popular convolution techniques \cite{oord2016wavenet, wei2018revisiting, sandler2018mobilenetv2, chen2017deeplab}. 

% \begin{figure}[hbt!]
%   \centering
%    \includegraphics[width=0.5\linewidth]{dilated.drawio.pdf}

%    \caption{
%    Difference between Traditional and Dilated Convolution \cite{holschneider1990real}: in this case, the dilation rate is set = 3 i.e there is 2 spaces inserted between kernel elements}
%    \label{fig:dilatedconv}
% \end{figure}

The dilated convolution expands the kernel by introducing gaps between its elements, effectively “inflating” it. The dilation rate, an additional parameter, determines the extent of expansion or widening of the kernel. From the formulation for traditional convolution:
 \begin{equation}
  (F*k)(p) = \sum_{s+t=p}F(s)k(t).
  \label{eq:important}
\end{equation}
The dilated convolution is determined as:
 \begin{equation}
  (F*_lk)(p) = \sum_{s+lt=p}F(s)k(t).
  \label{eq:important}
\end{equation}
With \(F\) is the input, \(k\) is the the kernel, \(s\), \(t\) is the position of the considered elements, \(l\) is the dilation rate.
% Figure \ref{fig:dilatedconv} visualizes the difference between the traditional convolution and dilated convolution. 

\subsubsection{Frame-wise regressor}
Regression on time-series data requires the sequence model to operate. Due to the effectiveness of Transformer in time-series processing tasks \cite{liu2021pyraformer, zhou2022fedformer, wu2021autoformer, zhou2021informer}, we decide to use it for our regression module. The transformer layers adopt a unique strategy by modeling pairwise interactions among temporal tokens within each layer. This design enables the transformer layer to effectively capture long-range dependencies throughout the entire time series sequence, starting from the initial layer. 
% Using transformer after Dilated Convolution help the model efficiently and effectively captures temporal information present in the input data, leading to high-performance outcomes while maintaining real-time processing capabilities.
Given an extracted time-series embedding from convolution layers, we employ Position Embedding in conjunction with these features to form the order for the sequence of tokens. Subsequently, these tokens are input into Transformer layers, consisting of Multi-Headed Self-Attention (MSA) \cite{vaswani2017attention}, layer normalization (LN) \cite{ba2016layer}, and MLP blocks.

% \begin{equation}
%   y^l = MSA(LN(z^l)) + z^l
%   \label{eq:important}
% \end{equation}
% %-------------------------------------------------------------------------
% \begin{equation}
%   z^{ l+1} = MLP(LN(y^l)) + y^l
%   \label{eq:important}
% \end{equation}

Given that the Transformer was originally designed for translation problem, we make slight modifications to adapt the model for multi-label classification tasks. Specifically, we treat the transformer as an auto-encoder to generate the sequence of meaningful information. Then, a fully-connected layer will receive this information to return the engagement score for each frame.
\begin{figure*}
  \centering
   \includegraphics[width=.90\linewidth]{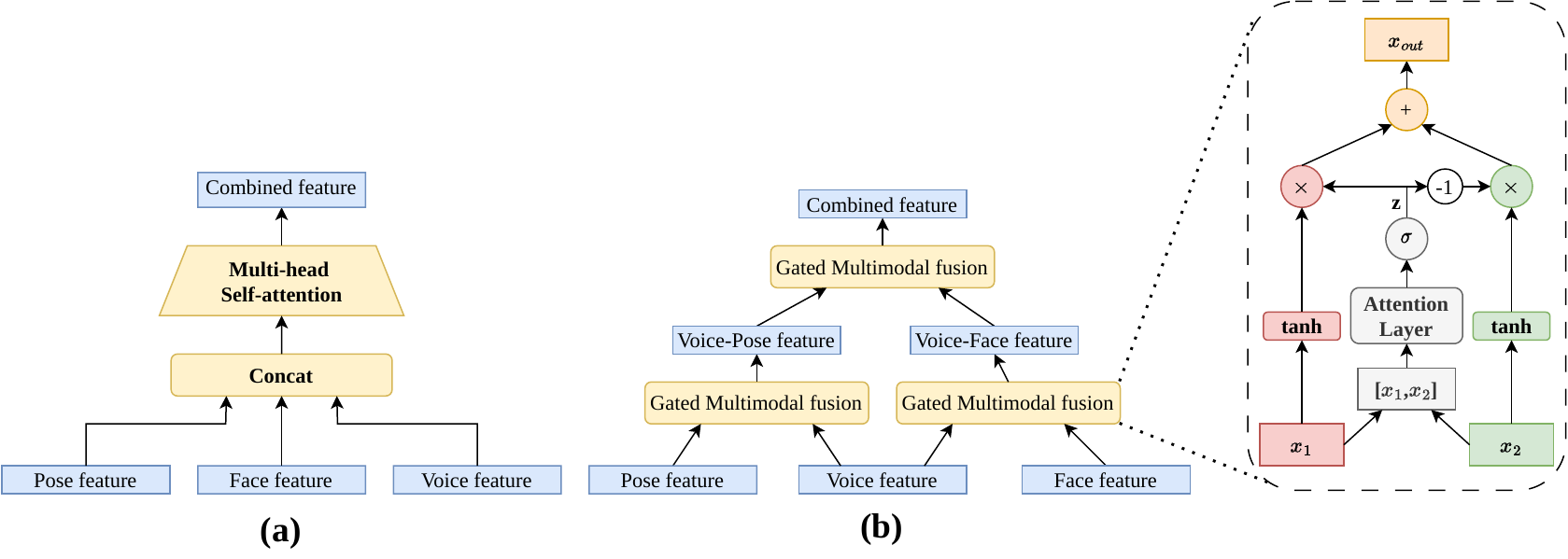}
   \caption{
   Two types of fusion strategies considered in our work: (a) Self-attention fusion, (b) Multimodal Gated Fusion.}
   \label{fig:fusion}
\end{figure*}

\subsection{Modalities fusion}
We employ two different fusion methods to find the good strategy of combining modalities information together. These fusion methods are desmonstrated in Figure \ref{fig:fusion}.
% \begin{figure*}
% \caption{Two multimodal combination strategy}
%      \centering
%      \begin{subfigure}[b]{0.3\textwidth}
%          \centering
%          \includegraphics[width=\textwidth]{self_attention_fusion.drawio.pdf}
%          \caption{Self-attention fusion}
%          \label{fig:y equals x}
%      \end{subfigure}
%      \hfill
%      \begin{subfigure}[b]{0.5\textwidth}
%          \centering
%          \includegraphics{Modality combination.drawio.pdf}
%          \caption{Multimodal gated fusion}
%          \label{fig:three sin x}
%      \end{subfigure}
% \end{figure}

\subsubsection{Self-attention fusion}
From the feature extracting from convolution layers, we use the naive channel-wise concatenation to merge the modalities \cite{nagrani2021attention, zhu2020multimodal}. Each frame feature is considered as an token and is concatenated right before passing to the transformer model. 
Despite being a simple strategy, this method already has already been proven as effective in different types of multimodality models. Feeding whole feature without fusions and alterations allow the attention layers to fuse the information itself. Hence utilize better the robustness of attention layers in mixing and finding the most informative components.
\subsubsection{Multimodal Gated Fusion}
The Gated Multimodal Unit (GMU) \cite{chen2020multi, saeed2022fusion, arevalo2017gated,saeed2022learning} is a model that draws inspiration from flow control mechanisms found in recurrent architectures such as GRU or LSTM. The GMU is designed to serve as an internal unit within a neural network architecture, aiming to generate an intermediate representation by combining data from different modalities.

In the GMU the feature vectors associated with each modality denoted as \(x_1\), \(x_2\) are fed into neurons with a tanh activation function, which encode internal representation features based on their respective modalities. For every input modality, there exists a gate neuron (Multiplication node) responsible for controlling the contribution of the feature derived from input feature vector to the overall output of the unit. This gated neuron serves as an attention layer, analyzing inter-modality relationships to determine the relevance of each modality in encoding a specific input sample. To fuse all three modalities, a hierarchical architecture is constructed.

% In this case, we construct a hierarchical architecture to fuse all three modals. 
% \subsubsection{Learning Participants Interaction }
\section{Experiments and Results}
\subsection{Datasets}
Engagement Estimation task in MULTIMEDIATE challenge employ Novice Expert Interaction (NOXI) Dataset \cite{cafaro17_icmi} as the benchmark. This dataset contains 128 videos and audio files: 76 training, 20 validation and 32 testing recording 64 conversation session between two participants. In these conversations, one participant is assumed as an expert and the other participant plays the role of a novice. In total, there are 2502433 annotated frames in the training and validation set of the dataset. NoXI dataset also provides signals recorded from those session in terms of three modalities: head, pose, and voice. 
\subsection{Experiments Settings}
All experiments were conducted on a GTX 3090 GPU using PyTorch Lightning for the implementation of the entire pipeline. The model was trained with the Adam optimizer, employing a learning rate of 1e-6, over 30 epochs. The input sequence size was set to 64 frames.
For the Dilated Convolution, we utilized three 1D Convolution Layers with kernel sizes of 5, 5, and 3, respectively, and a dilation rate of 4. In terms of the Transformer architecture, we employed a full model with 4 encoder layers and 4 decoder layers. The self-attention layers had 8 heads, with a hidden size of 128.

The evaluation metric used for the Engagement Estimation task of NOXI dataset is the Concordance Correlation Coefficient (CCC). We also use this metrics for loss function. CCC is formulated as:
\begin{equation}
\rho_{c} = \frac{(2\rho\sigma_x\sigma_y)}{\sigma_x^2+\sigma_y^2+(\mu_x-\mu_y)^2} 
\end{equation}
 where 
\(\mu _{x}\) and 
\(\mu _{y}\) are the means for the two variables and \(\sigma_{x}^{2}\) and 
\(\sigma_{y}^{2}\) are the corresponding variances. 
\(\rho\) is the correlation coefficient between the two variables.

\subsection{Experimental Results}
\subsubsection{Fusion strategy, Sequence model and loss function comparison}

% Table \ref{tab: overall} show our intensive experiment in different config and setting. 
% The report scored on validation set is the best validation score of the model and test score obtained from challenge portal. From the table, we can see that Dilated Convolution along with Self-attention fusion achieve best performance in the Test set at 0.66. The score of Gated Fusion with Transformer despite of gaining the best score on validation set (0.77) is drop significantly in the test set (0.6). This can be expressed as the over-fitting situation.
Table \ref{tab: overall} presents the results of our comprehensive experiments conducted with various configurations and settings. The reported scores on the validation set represent the best validation score achieved by the model. 

Based on the analysis of the table, it is evident that the combination of Dilated Convolution and Self-attention fusion achieves the highest performance on the test set, with a score of 0.66. However, the Gated Fusion with Transformer model, despite obtaining the best score on the validation set (0.77), experiences a significant drop in performance on the test set (0.6), showing the occurrence of overfitting. Furthermore, the unsatisfactory results obtained when training and validating each role independently (Expert: 0.58, Novice: 0.61) suggest that employing separate pipelines for each participant's role may not be an effective choice for this model, despite its initial intuitiveness.
\begin{table}[h]
\caption{Experiment result on different setting: different type of modules and participants.}
\begin{tabular}{|l|l|l|l|l|l|}
\hline
Convolution & Fusion & Regressor & Subject & Val  & Test \\ \hline
Dilated     & SA     & Transformer    & Ex      & 0.62 & \_   \\
Dilated     & SA     & Transformer    & No      & 0.65 & \_   \\
No          & SA     & Transformer    & Ex+No   & 0.67 & \_   \\
Traditional & SA     & Transformer    & Ex+No   & 0.70 & \_   \\
Dilated     & SA     & Cross-val LSTM & Ex+No   & 0.73 & 0.53 \\
Dilated     & GF     & LSTM           & Ex+No   & 0.72 & 0.53 \\
Dilated     & SA     & LSTM           & Ex+No   & 0.75 & 0.63 \\
Dilated     & GF     & Transformer    & Ex+No   & \textbf{0.77} & 0.60 \\
Dilated     & SA     & Transformer    & Ex+No   & 0.75 & \textbf{0.66} \\ \hline
\end{tabular}
\label{tab: overall}
 (SA: Self-attention; GF: Gated Fusion)
 (Ex: Expert; No: Novice)
\end{table}
\begin{table}[]
\caption{Result of our methods comparing with challenge baseline and other competitors}
\begin{tabular}{|l|l|l|l|}
\hline
                & Feature           & Val  & Test \\ \hline
Head    \cite{muller2023multimediate}         & AUs               & 0.31 & 0.22 \\
Body    \cite{muller2023multimediate}         & Openpose          & 0.54 & 0.44 \\
Voice   \cite{muller2023multimediate}         & gemaps            & 0.58 & 0.55 \\
Baseline \cite{muller2023multimediate}       & All features & 0.71 & 0.59 \\
USTC-IAT-United & \_                & \_   & 0.71 \\
Our             & All features & 0.75 & 0.66 \\ \hline
\end{tabular}
\label{tab: leaderboard}
\end{table}
\subsubsection{Comparison with the baseline and competitors on Leaderboard}
The leaderboard presented in Table \ref{tab: leaderboard} demonstrates the robustness of our model compared to the baseline, showcasing a 7\% improvement on the test set and a 4\% improvement on the validation set. However, our results still lag behind the top-ranked team in the challenge by a difference of 5\% on the test set. It is worth noting that although there were other teams submitting to the leaderboard, we only consider the team that indicated their intention to submit a paper.

% Furthermore, since the top-ranked competitor has not published their paper, it remains unknown whether they utilized the same or different types of features. This lack of information hinders a direct comparison of our approach with theirs.
\subsubsection{Ablation Study: Validating modalities contribution on Engagement in Conversation}
% In this subsection, our objective is to conduct a comprehensive examination of the contributions made by various modalities towards engagement within a conversation. While the preliminary findings have already established the significance of speech in indicating attentiveness, we aim to undertake a more thorough investigation and directly compare the correlation between the engagement score and each individual modality. In order to assess the correlation between two variables, we employ the CCC score to compute the correlation between each modality and the ground truth in the training set.

% Additionally, considering the multiple dimensions within each frame of the features, we employ the L2 norm to evaluate the feature magnitude. This allows for quantifying the magnitude of individual features and facilitates correlation computations. Furthermore, we conduct separate model training on each modality to conduct a precise comparison and identify the characteristics of each modality. The results of this analysis are presented in Table \ref{tab: ablation}.
% In this subsection, we analyze the impact of different modalities on conversation engagement. Baseline results highlight the crucial role of speech in indicating attentiveness. Using the CCC score and feature magnitude, we assess the correlations between modalities and engagement. 
In this subsection, we analyze the impact of different modalities on conversation engagement. We delve deeper to directly compare engagement score correlations with each modality. Using the CCC score and feature magnitude, we assess the correlations between modalities and engagement. This analysis provides insights into modality importance and correlations with engagement. Results are presented in Table \ref{tab: ablation}. It reveals an interesting contrast in our analysis. While speech remains crucial in predicting engagement, there is confusion regarding the importance of the Head and Pose modalities. Initially, the baseline scores suggested the Pose modality's significance over the Head modality. However, considering the correlation between Feature Magnitude and our model scores, the opposite trend emerges. 

\begin{figure}
  \centering
   \includegraphics[width=.95\linewidth]{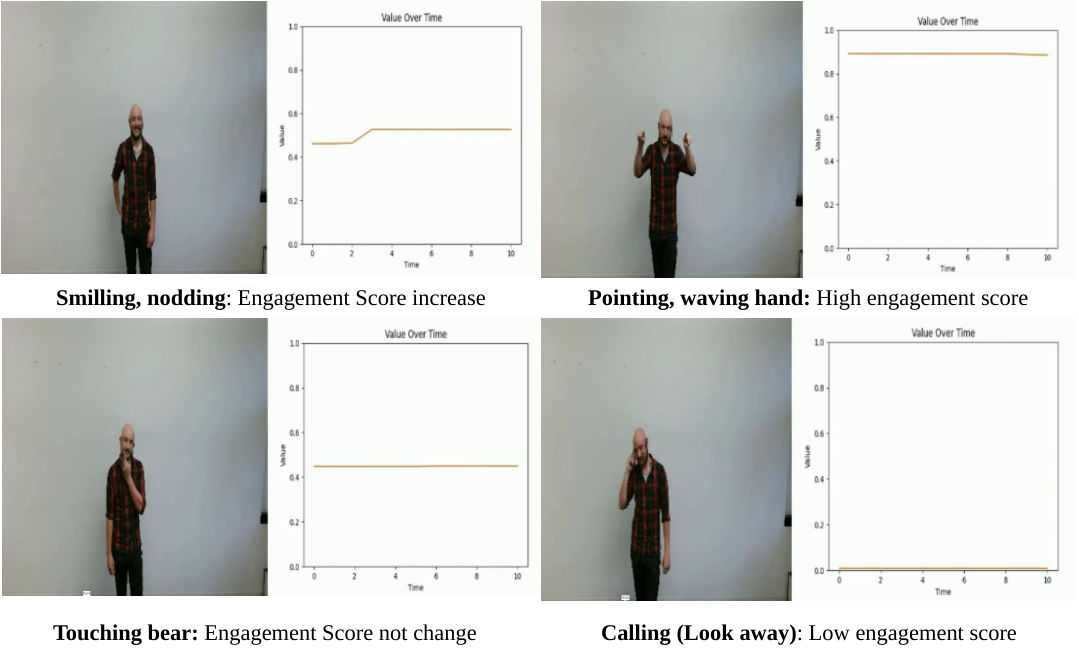}
   \caption{
   Sample from the novice video in session 69. The variation in the Pose and Facial expression in context of the participants posed a huge challenge for engagement estimation in conversation.}
   \label{fig:sample}
\end{figure}

To gain further insights, we visualize some sample data in Figure \ref{fig:sample}, demonstrating the complexities arising from variations in human pose and facial expressions in different contexts. Smiling while listening to opponents increases the engagement score, while smiling during a call does not. Similarly, actions like leaning the head and waving the hand to touch the beard have no impact on engagement. However, waving the hand to point indicates full engagement in the conversation. These context-dependent variations pose challenges for accurate engagement estimation. 
\begin{table}[]
\caption{Comparing CCC between Ground truth and Feature Magnitude with models' score.}
\begin{tabular}{|l|l|l|l|}
\hline
                & Feature magnitude & Baseline Score  & Our Score \\ \hline
Head      & 0.0031            & 0.31           & 0.435     \\ \hline
Pose  & 0.0003            & 0.54           & 0.192     \\ \hline
Voice   & 0.0069            & 0.58           & 0.59      \\ \hline
\end{tabular}
\label{tab: ablation}
\end{table}

\section{Conclusion}
Our paper presents dilated convolution based Transformer model for Engagement Estimation in the MULTIMEDIATE Competition. It outperforms the baseline by incorporating dilated convolution and transformer layers, achieving better long-term capture ability. We also find that the Self-attention fusion strategy yields the best results among two multimodal fusion approaches. However, our method shows overfitting with decent validation results but lower test set performance, indicating the need to address this for better generalization.

% This 
\begin{acks}
 This work was supported by the National Research Foundation of Korea (NRF) grant funded by the Korea government (MSIT) (RS-2023-00219107). This work was also supported by Institute of Information \& communications Technology Planning \& Evaluation (IITP) grant funded by the Korea government(MSIT) (No.2021-0-02068, Artificial Intelligence Innovation Hub), and the Artificial Intelligence Convergence Innovation Human Resources Development (IITP-2023-RS-2023-00256629) grant funded by the Korea government (MSIT).
\end{acks}

\bibliographystyle{ACM-Reference-Format}
\bibliography{refs}

\end{document}